\documentclass[journal]{IEEEtran}
\usepackage{multirow} 
\usepackage{bbding}
\usepackage{makecell}
\usepackage{color}
\usepackage{amssymb}
\usepackage{amsmath,graphicx,cite}
\usepackage{comment}
\usepackage{subcaption}
\usepackage{tabularx}
\usepackage{diagbox}

\newcolumntype{R}[1]{>{\raggedleft\arraybackslash}p{#1}}
\newcolumntype{C}[1]{>{\centering\arraybackslash}p{#1}}
\newcolumntype{L}[1]{>{\raggedright\arraybackslash}p{#1}}
\ifCLASSINFOpdf
\else
\fi

\begin{document}
\title{Cross-subject dual-domain fusion network with task-related and task-discriminant component analysis enhancing one-shot SSVEP classification }
\author{Yang~Deng, Zhiwei Ji, Yijun~Wang,~\IEEEmembership{Member,~IEEE,}     and~S.Kevin~Zhou,~\IEEEmembership{Fellow,~IEEE}
\thanks{This work was supported by the National Key R\&D Program of China under grant 2022YFF1202303, the National Natural Science Foundation of China under grant 62071447, and the Strategic Priority Research Program of Chinese Academy of Sciences under grant XDB32040200. (Corresponding authors: Yijun Wang; S. Kevin Zhou.)}
\thanks{Yang Deng and S. Kevin Zhou are with the School of Biomedical Engineering, Division of Life Sciences and Medicine, University of Science and Technology of China, Hefei, Anhui, and also with the Center for Medical Imaging, Robotics, Analytic Computing \& Learning (MIRACLE), Suzhou Institute for
Advanced Research, University of Science and Technology of China, Suzhou, Jiangsu, China}
\thanks{Zhiwei Ji is with the College of Artificial Intelligence, Nanjing Agricultural University, Nanjing 210095, China}
\thanks{Yijun Wang is with the State Key Laboratory on Integrated Optoelectronics, Institute of Semiconductors, Chinese Academy of Sciences,  Beijing, China}
}

\markboth{Journal of \LaTeX\ Class Files,~Vol.~14, No.~8, August~2015}%
{Shell \MakeLowercase{\textit{et al.}}: Bare Demo of IEEEtran.cls for IEEE Journals}

\maketitle

\begin{abstract}
This study addresses the significant challenge of developing efficient decoding algorithms for classifying steady-state visual evoked potentials (SSVEPs) in scenarios characterized by extreme scarcity of calibration data, where only one calibration is available for each stimulus target. To tackle this problem, we introduce a novel cross-subject dual-domain fusion network (CSDuDoFN) incorporating task-related and task-discriminant component analysis (TRCA and TDCA) for one-shot SSVEP classification. The CSDuDoFN framework is designed to comprehensively transfer information from source subjects, while TRCA and TDCA are employed to exploit the single available calibration of the target subject. Specifically, we develop multi-reference least-squares transformation (MLST) to map data from both source subjects and the target subject into the domain of sine-cosine templates, thereby mitigating inter-individual variability and benefiting transfer learning. Subsequently, the transformed data in the sine-cosine templates domain and the original domain data are separately utilized to train a convolutional neural network (CNN) model, with the adequate fusion of their feature maps occurring at distinct network layers. To further capitalize on the calibration of the target subject, source aliasing matrix estimation (SAME) data augmentation is incorporated into the training process of the ensemble TRCA (eTRCA) and TDCA models. Ultimately, the outputs of the CSDuDoFN, eTRCA, and TDCA are combined for SSVEP classification. The effectiveness of our proposed approach is comprehensively evaluated on three publicly available SSVEP datasets, achieving the best performance on two datasets and competitive performance on one. This underscores the potential for integrating brain-computer interface (BCI) into daily life. The corresponding source code is accessible at https://github.com/Sungden/One-shot-SSVEP-classification.
\end{abstract}

\IEEEpeerreviewmaketitle

\section{Introduction}
\label{sec:intro}
\IEEEPARstart{S}{teady}-state visual evoked potential-based brain-computer interfaces (SSVEP-BCI) have garnered considerable attention due to their simplicity, minimal training data requirements, and the attainment of high information transfer rates(ITR)\cite{Abiri_Borhani_Sellers_Jiang_Zhao_2019}\cite{wang2006practical}. Ongoing research in SSVEP-BCI decoding methods is broadly classified into two primary paradigms: knowledge-driven and data-driven, further subdivided into calibration-based and calibration-free based on the utilization of calibration data. In the knowledge-driven realm, canonical correlation analysis (CCA)\cite{bin2009online}\cite{lin2006frequency} is a prominent calibration-free approach, with its primary objective being the identification of two spatial filters optimizing the correlation coefficient between the filtered test signal and the filtered sine-cosine templates. In the domain of calibration-based methods, task-related component analysis (TRCA) \cite{nakanishi2017enhancing} and task-discriminative component analysis (TDCA) \cite{liu2021improving} algorithms have exhibited remarkable performance. TRCA focuses on training a spatial filter that maximizes the cumulative correlation coefficients among filtered trials, emphasizing response repeatability across trials, whereas the TDCA algorithm employs a discriminative model to identify a common spatial filter for all stimulus frequencies, thereby reducing the spatial filter redundancy of TRCA. To effectively harness the harmonic components inherent in SSVEP signals, filter bank (FB) techniques have been devised and widely integrated into CCA, TRCA, and TDCA methods, resulting in FBCCA\cite{chen2015filter}, FBTRCA\cite{nakanishi2017enhancing} and FBTDCA \cite{liu2021improving}. In the realm of data-driven approaches, artificial neural networks (ANNs), propelled by advancements in deep learning technology, have found utility in SSVEP classification. These data-driven methods can directly exploit data from diverse subjects and are applicable in both calibration-based and calibration-free scenarios. Among these, notable methodologies include the compact convolutional neural network (Compact-CNN) \cite{waytowich2018compact}, Conv-CA \cite{li2020convolutional}, bi-SiamCA\cite{zhang2022bidirectional}, deep neural network (DNN) \cite{guney2021deep} and TRCA-Net \cite{deng2023trca}. 

While calibration-based algorithms, such as TRCA, TDCA, and deep learning-based approaches, have demonstrated exceptional performance in SSVEP decoding, they often require a substantial amount of individual data for effective training. They can face limitations when dealing with limited calibration data. For instance, due to the instability of the estimated covariance matrix used in spatial filter generation and the low signal-to-noise ratio (SNR) of averaged SSVEP templates across calibration trials \cite{wong2020learning} \cite{luo2022data}, the accuracy of TRCA and TDCA approaches tends to deteriorate or even become invalid under these conditions. In the case of deep learning-based approaches, the presence of limited training data can lead to overfitting \cite{srivastava2014dropout}. However, collecting multiple EEG trials for SSVEP-BCIs can be time-consuming and labor-intensive, especially for systems with numerous targets. To address the challenge of limited data, advanced calibration-free approaches have been proposed. For example, Guney et al. \cite{guney2023transfer} introduced an ensemble DNN approach, which entails training a global target identifier DNN using source datasets, followed by fine-tuning for each individual target data. Chen et al. \cite{chen2023transformer} introduced a transformer-based approach that utilized the frequency spectrum of SSVEP data as input and explored both spectral and spatial domain information for classification. Huang et al. \cite{huang2023cross} proposed a cross-subject transfer method based on domain generalization, which transfers domain-invariant spatial filters and templates learned from source subjects to the target subject. Although calibration-free methods have shown some promise, there remain several challenges that demand attention. Notably, there is a considerable scope for enhancing decoding accuracy. For instance, in the context of the Benchmark dataset \cite{wang2016benchmark} and the BETA dataset \cite{liu2020beta}, the ensemble DNN \cite{guney2023transfer}, considered the state-of-the-art (SOTA) calibration-free approach, achieves accuracy levels of only 53.92 and 47.61 at 0.5 seconds, in contrast to the notably higher accuracy of the calibration-based ensemble TRCA (eTRCA) \cite{nakanishi2017enhancing} at 77.55 and 59.21. 

As a bridging approach, the less and minimal calibration method has gained prominence and demonstrates substantial potential. Chiang et al. \cite{chiang2021boosting} introduced a transfer learning approach based on least-squares transformation (LST), leveraging calibration data across multiple domains, including sessions, subjects, and EEG montages. Liu et al. \cite{liu2021align} introduced a recalibration-free cross-device transfer learning framework, referred to as ALign and Pool for EEG Headset domain Adaptation (ALPHA),  boost the performance of dry electrode-based SSVEP-BCI. Wong et al. \cite{wong2020learning} proposed the multi-stimulus learning algorithm, which can learn data corresponding to not only the target stimulus but also other stimuli. Moreover, Wong et al. \cite{wong2020inter} introduced a subject transfer-based canonical correlation analysis (stCCA) algorithm, utilizing knowledge within and between subjects, which requires only a small amount of calibration data from a new subject. Recently, Luo et al. \cite{luo2022data} introduced a data augmentation approach known as source aliasing matrix estimation (SAME), which utilizes sine-cosine templates to enhance the performance of TRCA and TDCA, especially in scenarios with limited calibration data. Among these methods, stCCA stands out as the most promising and holds the SOTA status \cite{wong2020inter}. However, it's crucial to acknowledge that the second assumption proposed in stCCA \cite{wong2020inter}, suggesting the sharing of knowledge among spatially filtered SSVEP templates from different subjects and using a weighted summation
of spatially filtered SSVEP templates from source subjects
to approximate the spatially filtered SSVEP template of the
target subject lacks robust supporting evidence. Concerning recently introduced minimal calibration approaches, it's essential to highlight that some of them, including generalized zero-shot learning (GZSL) \cite{wang2023generalized} and periodically repeated component analysis (PRCA) \cite{ke2023enhancing}, may not have been thoroughly compared to the stCCA approach.
Furthermore, reported results may not entirely reflect the actual performance achieved by the SOTA method, as evidenced by small data least-squares transformation (sd-LST) \cite{bian2022small}. For instance, the ITR for stCCA on the Benchmark dataset for the 0.7-second signal length with $N_{trial}$ set to 9 was reported to be over 198 \cite{wong2020inter}. However, the result of stCCA reported in sd-LST \cite{bian2022small} was 186 with $N_{trial}$ set to 15, significantly lower than the original SOTA results.

This study proposes a novel cross-subject dual-domain fusion network (CSDuDoFN) with eTRCA and TDCA to enhance one-shot SSVEP classification. Our method is motivated to achieve the following two primary objectives: (1) to comprehensively mitigate the domain gap between the source and target data, ultimately enhancing the effectiveness of transfer learning on the target data; (2) to make optimal use of the single calibration data from the target subject. We develop the CSDuDoFN architecture, which can transform source subjects and target subjects into a unified domain, narrowing their domain gap for better transfer learning. Specifically, we apply multi-reference LST (MLST) on each individual SSVEP trial with each stimulus, generating new multi-channel signals. Subsequently, the transformed data and the original domain data are separately utilized for training a convolutional neural network (CNN) model, with the fusion of their feature maps occurring at distinct network layers. Moreover, to further capitalize on the target data calibration, source aliasing matrix estimation (SAME) data augmentation \cite{luo2022data} is incorporated into the training process of the eTRCA and TDCA models. Ultimately, the output of the CSDuDoFN, eTRCA, and TDCA are combined for the final SSVEP classification. Notably, in our approach, we have incorporated both eTRCA and TDCA instead of exclusively relying on TDCA. This decision is based on the understanding that TDCA does not consistently outperform eTRCA. Previous research has indicated that in some cases, eTRCA has demonstrated superior performance to TDCA \cite{luo2022data} \cite{xiao2022fixed}. By using both eTRCA and TDCA, we aim to capitalize on their complementary strengths. To assess the effectiveness of our approach, we carry out experiments using three well-known publicly available datasets: the UCSD dataset \cite{nakanishi2015comparison}, the Benchmark dataset \cite{wang2016benchmark}, and the BETA dataset \cite{liu2020beta}. The outcomes of these experiments clearly indicate that our proposed method attains promising results for one-shot SSVEP classification.

The paper is structured as follows: Section \ref{sec:pre} provides the necessary preliminaries. Section \ref{sec:method} describes the proposed approach. The experiments and results are presented in Section \ref{sec:results}. The discussion and conclusion are given in Sections \ref{sec:Dis} and\ref{sec:Con}, respectively. 

\section{Preliminaries}\label{sec:pre}
This section introduces the problem definition, notions used in this paper, and preliminary algorithms, including the SAME data augmentation method, TRCA, and TDCA.

\subsection{Problem Definition and Notions}
Constrained by calibration time, in many cases in SSVEP-BCI systems, there is only one calibration data available for each visual stimulus, i.e., $K$ = $N_f$. We refer to the SSVEP decoding in this scenario as "one-shot SSVEP classification." While some methods can operate when $K \leq N_f$, such as the stCCA method \cite{wong2020inter}, sd-LST \cite{bian2022small} and GZSL \cite{wang2023generalized}, in this paper, we compare their results when $K$ = $N_f$. For a clear introduction to the SSVEP recognition algorithms, Table.~\ref{Notions} summarizes the notations that are used in this paper. 

\begin{table}[htbp]
\caption{Table of notations}
\label{Notions}
\begin{tabular}{|m{3cm}|m{5cm}|}
    \hline
    \textbf{Notation} & \textbf{Description} \\
    \hline
    $N_f$  & Number of visual stimuli. \\\hline
    $N_c$  & Number of channels. \\\hline
    $N_s$  & Number of sampling points. \\\hline
    $N_h$  & Number of harmonics. \\\hline
    $f_s$  & Sampling rate. \\\hline
    $f_t$  & Frequency of the $t$-th visual stimulus. \\\hline
    $\phi_t$  & Phase of the $t$-th visual stimulus. \\\hline
    $K$  & Number of visual stimuli from the target subject. \\\hline
    $N_{trial}$  & Number of training trials for all stimuli. \\\hline
    $\mathbf{X}_{t}^{(i)}\in \mathbf{R}^{{N_c} \times {N_s}}$  & The $i$-th single-trial multi-channel SSVEP signal for $t$-th stimulus. \\\hline
    $\mathbf{X}_{tt}^{(i)} \in \mathbf{R}^{{2N_h} \times {N_s}}$ & The $i$-th transformed signal corresponding to LST for $t$-th stimulus. \\\hline
    $\mathbf{\overline{X}}_{t}\in \mathbf{R}^{{N_c} \times {N_s}}$  & SSVEP template corresponding to the $t$-th stimulus. \\\hline
    $\mathbf{Y_t}\in \mathbf{R}^{{2N_h} \times {N_s}}$ & The sine-cosine template for $t$-th stimulus. \\\hline
    $\mathbf{P}_{LST} \in \mathbf{R}^{{2N_h} \times {N_c}}$ & LST matrix from labeled one-shot data.\\\hline
\end{tabular}
\end{table}

\subsection{SAME Data Augmentation}
SAME data augmentation \cite{luo2022data} is designed to use sine-cosine templates for generating artificial signals. To achieve this, the algorithm first estimates the LST matrix that relates the target signals to the sine-cosine templates. Subsequently, this estimated matrix is applied to the sine-cosine templates, and random Gaussian noise is added to create the artificial signals. When combined with TRCA and TDCA, this method significantly improves performance, even when only a small amount of calibration data is available. This approach has the potential to substantially minimize the calibration efforts, which holds promise for the advancement of practical BCIs. Please consult the original study in \cite{luo2022data} for more detailed information.
\subsection{TRCA and TDCA}
The core concept of TRCA is to extract task-related signal components from observed SSVEP signals and create spatial filters by maximizing inter-trial covariance. However, prior research \cite{wong2020learning} in SSVEP-BCI has revealed that different categories of stimuli tend to share a common spatial pattern or spatial filter. Consequently, TRCA typically employs ensemble techniques (eTRCA) to enhance performance by connecting spatial filters for each category \cite{nakanishi2017enhancing}. Nevertheless, this cascade of spatial filters introduces redundancy, and so far, there hasn't been a well-understood mechanism for determining the optimal level of redundancy. Given the similarity in spatial patterns among different stimuli, it's unnecessary to learn spatial filters separately for each stimulus. As a result, TDCA recommends learning the projection direction shared by all data categories \cite{liu2021improving}. Unlike TRCA, which relies on generative models, TDCA employs discriminant models to address frequency recognition challenges. Furthermore, the TDCA algorithm takes into consideration the possibility that time information in SSVEP may not be fully utilized. More details on TRCA and TDCA can be found in \cite{nakanishi2017enhancing} and \cite{liu2021improving}.

\section{Methods}\label{sec:method}
\subsection{Overview}
\begin{figure}[htbp]
    \centering
    \includegraphics[width=0.5\textwidth]{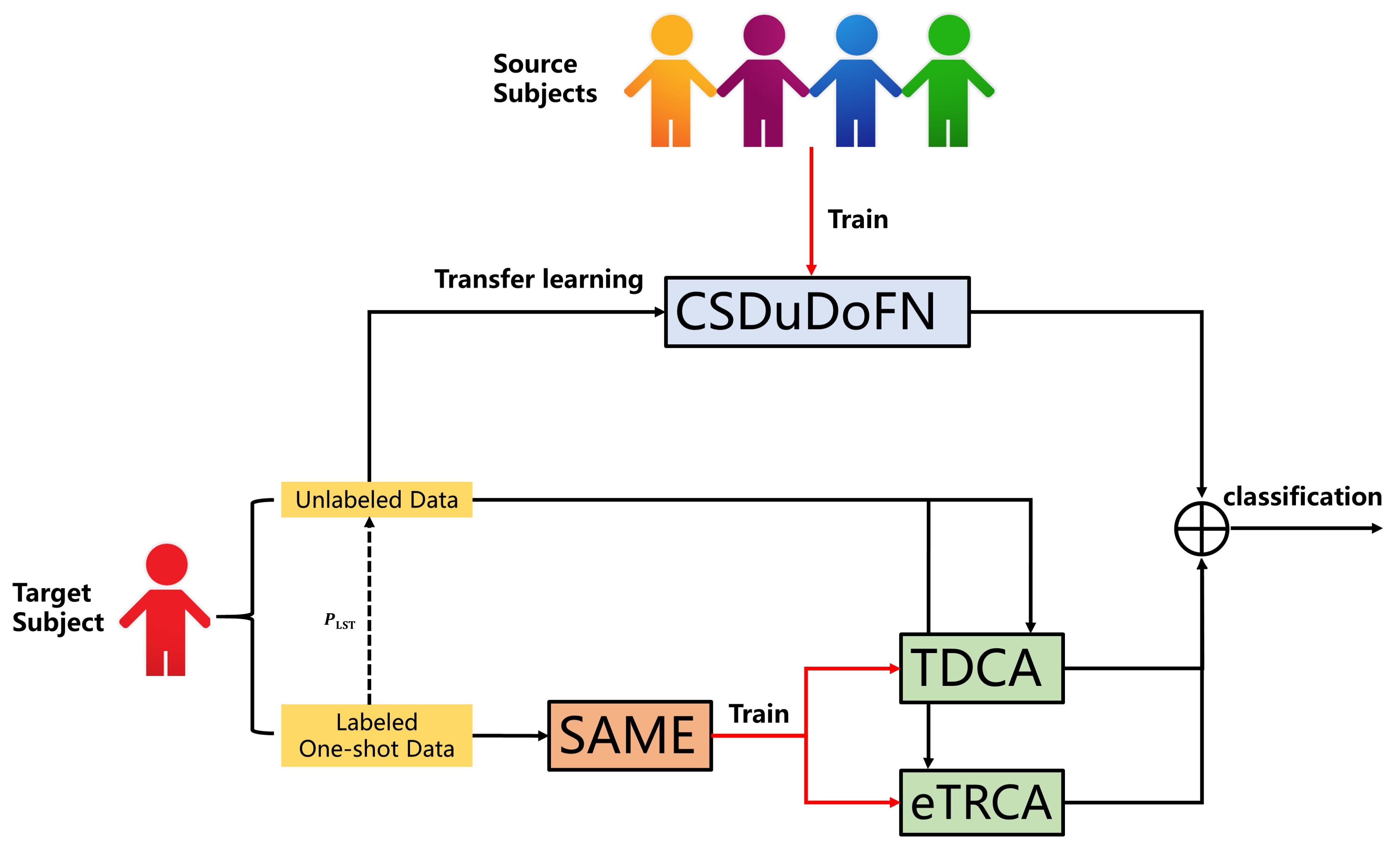}
    \caption{The overview of our proposed approach. The workflow begins by training the CSDuDoFN model using source subjects. Afterward, we apply SAME to the one-shot labeled data to train the TDCA and eTRCA models. Finally, the classification of unlabeled data from the target subject is achieved by combining the outputs of trained CSDuDoFN, eTRCA, and TDCA. }
    \label{fig:overview}
    \centering
    \vspace{-3mm}
\end{figure}

The overview of our approach is represented in Fig.~\ref{fig:overview}. We initiate the process by training the CSDuDoFN model using data collected from source subjects. For the new subject, who has only one calibration session for each stimulus, we employ SAME data augmentation to enhance the one-shot labeled data, facilitating the training of the TDCA and eTRCA models. Simultaneously, we compute the LST transformation matrices for each stimulus during the one-shot calibration and apply them to the unlabeled data. During the classification of this unlabeled data, it is fed into the pre-trained CSDuDoFN, TDCA, and eTRCA models. The final prediction is generated by combining the outputs of these three models. It's worth noting that since the one-shot data from the target subject is not used in the training of the CSDuDoFN model, the training time is minimal, making our approach highly suitable for online use.

\subsection{Multi-reference LST (MLST)}
LST, a commonly used statistical method, is employed to establish a linear relationship model between variables. Its primary goal is to estimate the parameters of the regression model by minimizing the sum of squared residuals, which quantifies the disparity between predicted values and actual observations. LST has found application in mitigating cross-subject variations by creating a transformation matrix between the target subject and each source subject for each stimulus frequency. Notably, Chiang et al. [1] and Luo et al. [2] independently utilized LST for SSVEP transfer learning and data augmentation, resulting in promising performance.

\begin{figure*}[t]
\centering
\includegraphics[width=1\textwidth]{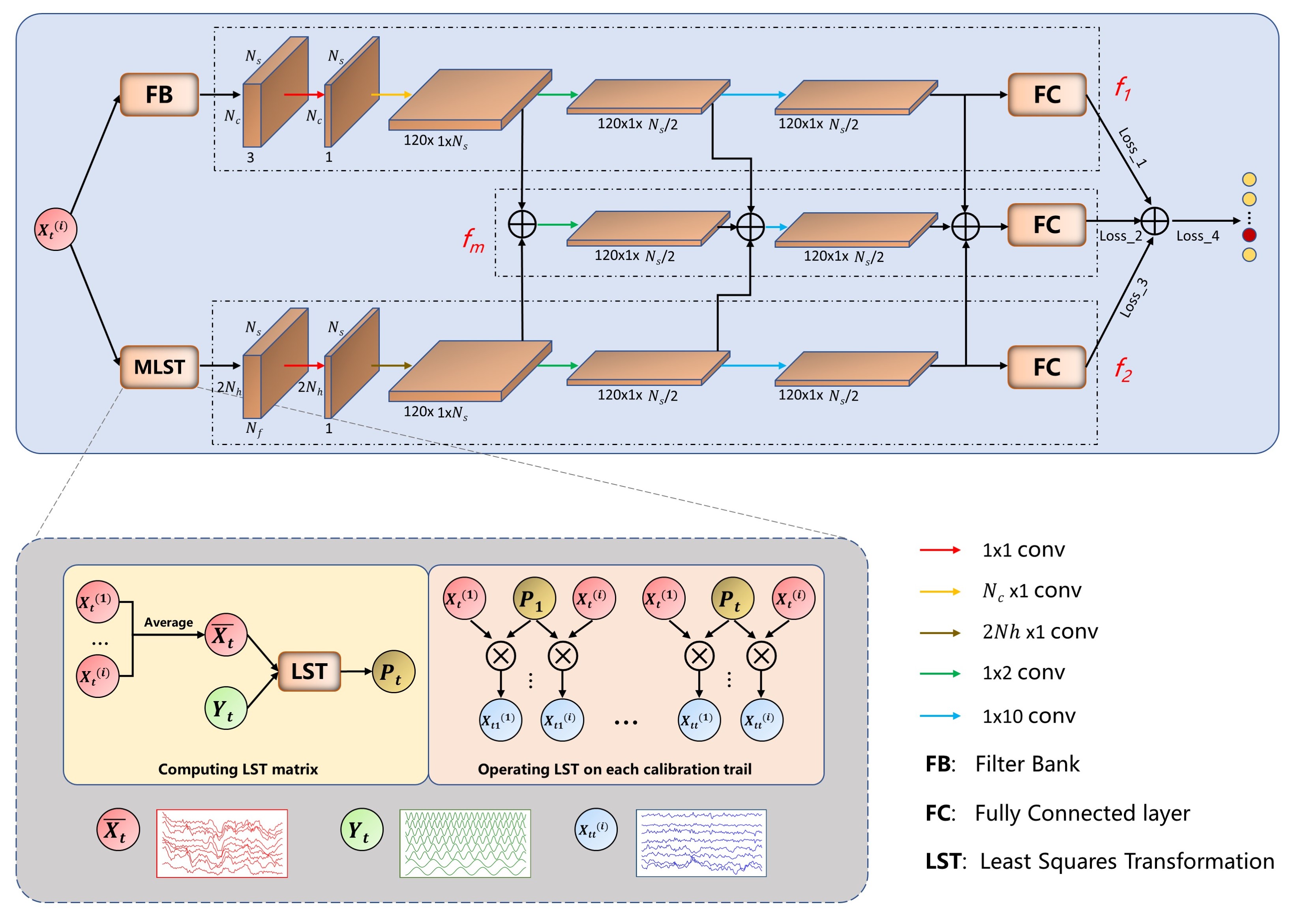}
\caption{The pipeline of our proposed CSDuDoFN. The original SSVEP signals are first subjected to filter bank decomposition before being fed into model $f_1$. After undergoing MLST processing, they are directed to model $f_2$. The intermediate features from models $f_1$ and $f_2$ are combined and sent to model $f_m$. The outputs of models $f_1$, $f_2$, and $f_m$ are summed to produce the final classification result. Additionally, their individual outputs, as well as the combined output, are used to calculate the loss.}
\label{fig:pipe}
\end{figure*}

When $\mathbf{X_1}$ and $\mathbf{X_2}$ denote two 2D variables, the goal of LST is to determine a transformation matrix $\mathbf{P}$ by solving the following equation:
\begin{eqnarray}
\mathbf{P}=\underset{\mathbf{P}}{\operatorname{argmin}}\left\|\mathbf{X_1}-\mathbf{P}\mathbf{X_2}\right\|_{2}^2 
\end{eqnarray}

The following equation provides the solution to this problem:
\begin{eqnarray}
\mathbf{P} &=& \mathbf{X_1}{\mathbf{X_2}}^T(\mathbf{X_2} \mathbf{X_2}^T)^{-1}
\end{eqnarray}

In this context, we apply Multi-reference LST (MLST) to each individual SSVEP trial with each stimulus, creating new multi-channel signals. Let $\mathbf{X}_{t}^{(i)}\in \mathbf{R}^{{N_c} \times {N_s}}$ represent the $i$-th single-trial SSVEP signal from the $t$-th stimulus, and $\mathbf{Y}_{t}$ denote the sine-cosine template. Our objective is to map SSVEP signals from different subjects to sine-cosine templates using the following equation:
\begin{eqnarray}
\mathbf{P_t}=\underset{\mathbf{P_t}}{\operatorname{argmin}}\left\|\mathbf{Y}_{t}-\mathbf{P_t}\mathbf{\overline{X}}_{t}\right\|_{2}^2, 
\end{eqnarray}
where$\overline{X}_{t}$ is the SSVEP template by averaging each trail.

For MLST, the sine-cosine template corresponding to the $t$-th stimulus is constructed according to equation \ref{ALST}. 
\begin{equation}
\label{ALST}
\mathbf{Y}_t=\left[\begin{array}{c}
\sin \left(2 \pi f_t t+\phi_t\right) \\
\cos \left(2 \pi f_t t+\phi_t\right) \\
\vdots \\
\sin \left(2 \pi N_h f_t t+N_h \phi_t\right) \\
\cos \left(2 \pi N_h f_t t+N_h \phi_t\right)
\end{array}\right], \boldsymbol{t}=\left[\frac{1}{f_s}, \frac{2}{f_s}, \ldots \frac{N_s}{f_s}\right],
\end{equation}
where $N_h$ is the number of harmonics, $f_t$ denotes the stimulus frequency, $\phi_t$ denotes the stimulus phase, $N_s$ represents the number of sampling points, and $f_s$ is the sampling rate. 

When $\mathbf{P}{_t}$ is obtained, we apply $\mathbf{P}{_t}$ $\times$ $\mathbf{X}^{(i)}$ to each single-trial SSVEP signal, resulting in the generation of transformed multi-channel signals. For example, in the case of the Benchmark dataset with 40 characters, there are 40 sine-cosine templates, each with a size of ${2N_h} \times {N_s}$, and after performing MLST on each trial, 40 transformed signals with a size of ${2N_h} \times {N_s}$ are generated.

\subsection{The Proposed CSDuDoFN}
The pipeline of our proposed CSDuDoFN model is illustrated in Fig.~\ref{fig:pipe}. This model comprises three networks: $f_1$, $f_2$, and $f_m$. Each network is responsible for extracting information from the SSVEP original domain, the MLST-transformed domain, and the fused information, respectively. Inspired by DNN\cite{guney2021deep}, both $f_1$ and $f_2$ consist of four convolutional layers. For the original SSVEP signals, we employ filter bank decomposition with three filters, breaking down the initial ${N_c} \times {N_s}$ signal into three separate ${N_c} \times {N_s}$ inputs for $f_1$. As for the MLST-transformed signals, we obtain ${2N_h} \times {N_c} \times {N_s}$ inputs for $f_2$. These inputs then pass through a 1$\times$1 filter convolutional layer to capture different harmonics. Subsequently, they are directed through 120 convolutional layers with filter sizes of $N_c\times$1 (for $f_1$) or 2$N_h\times$1 (for $f_2$) to extract information from different channels. Next, they traverse 120 convolutional layers with filter sizes of 1$\times$2 and 1$\times$10 to capture information at different time scales.

To further exploit information from the original and transformed domains, we combine the outputs of the second, third, and fourth convolutional layers from  $f_1$ and $f_2$, and apply convolution ($f_m$). Finally, the outputs of $f_1$, $f_2$, and $f_m$ are forwarded to fully connected layers for classification, and their individual classification results are summed to obtain the final classification result. It's essential to note that the settings, such as filter bank decomposition with three filters, convolution kernel sizes of 1$\times$2 and 1$\times$10, and 120 filters, draw inspiration from DNN\cite{guney2021deep}. To enhance the training process with more supervision, we introduce four loss functions, namely loss1, loss2, loss3, and loss4. These loss functions all follow a smooth cross-entropy format with a smoothing value of 0.01. The final loss function is computed as:

\begin{eqnarray}
loss &=& loss\_1 + loss\_2 + loss\_3 + loss\_4
\end{eqnarray}

\subsection{SSVEP Classification}
In line with the motivation presented in Section \ref{sec:intro}, which aims to maximize the utilization of single calibration data from the target subject, we have integrated SAME-augmented eTRCA and TDCA for the classification process. Following this, we normalize the output features from the eTRCA, TDCA, and CSDuDoFN models, scaling them to the [0, 1] range. The final classification result is determined by summing these normalized features and selecting the label associated with the highest value. Based on the results obtained from SAME \cite{luo2022data}, it was observed that data augmentation three times yields the best results when there is only one calibration data available. Therefore, in our experiments, we utilize the results of SAME after undergoing three augmentations. The SAME results we compare with have also been enhanced three times.

\section{EXPERIMENTS AND RESULTS}\label{sec:results}
\subsection{Datasets}
This study utilized SSVEP data from three publicly available SSVEP datasets: UCSD dataset \cite{nakanishi2015comparison}, Benchmark dataset \cite{wang2016benchmark}, and BETA dataset \cite{liu2020beta}. All three datasets are coded in a joint frequency and phase modulation (JFPM) approach.

1) \textbf{The UCSD dataset} This dataset features 8-channel EEG data recorded during a 12-class SSVEP experiment with 10 subjects. Stimuli frequencies range from 9.25 Hz to 14.75 Hz, with a 0.5 Hz interval. Phases vary from 0 to 2$\pi$, with a 0.5$\pi$ offset. Each participant completed 15 blocks, each with 12 trials. The dataset includes 1114 sampling points at a rate of 256 Hz, with visual stimulation onset at the 39th sample point, resulting in 0.15 seconds of redundant data before stimulus onset.

2) \textbf{The Benchmark dataset} This dataset includes 64-channel EEG data from 35 subjects (8 experienced, 27 naive) engaged in a cue-guided target-selection task. They used a virtual keyboard with 40 flickers, covering digits, letters, and symbols. Stimulation frequencies ranged from 8 Hz to 15.8 Hz with a 0.2 Hz interval and a 0.5$\pi$ phase difference between adjacent frequencies. Each subject completed six blocks of 40 trials, with flickers cued randomly, and each trial lasted five seconds. 

3) \textbf{The BETA dataset} Like the Benchmark dataset, BETA includes 40 targets and involves 70 participants. Notably, these experiments took place outside the lab, leading to a lower signal-to-noise ratio (SNR) and a more challenging target identification task. Stimulation duration is 2 seconds for the first 15 subjects and 3 seconds for the rest, with an average visual latency of approximately 130 ms for the subjects in this dataset.

In our experiments, we employ a classical montage to evaluate performance. For the Benchmark and BETA datasets, the montage includes Pz, PO3, PO5, PO4, PO6, POz, O1, Oz, and O2 channels. In the case of the UCSD dataset, the montage comprises PO7, PO3, POZ, PO4, PO8, O1, OZ, and O2 channels.

\subsection{Performance Evaluation}
The algorithm's performance is assessed using the metrics of accuracy and Information Transfer Rate (ITR) for various data lengths. ITR is measured in bits per minute (bpm) and is defined as \cite{wolpaw2002brain}:
{\small {
\begin{align}
\text{ITR} &= (\log_2M + P\log_2P + (1-P)\log_2 \left[ \frac{1-P}{M-1}\right])\frac{60}{T}  
\end{align}}}
\normalsize
Here, $M$ represents the number of classes, $P$ represents accuracy, and $T$ (in seconds) represents the selection time, including both gaze and shift time. A gaze shift time of 0.5 seconds is employed for analysis. The leave-one-out cross-validation approach is used to evaluate the effectiveness of the methods.

\subsection{Compared Methods}
We conduct a comparative analysis of our method against eight SOTA approaches: FBCCA \cite{chen2015filter}, Compact-CNN \cite{waytowich2018compact}, Ensemble DNN \cite{guney2023transfer}, eTRCA with SAME \cite{luo2022data}, TDCA with SAME \cite{luo2022data}, LST \cite{chiang2021boosting}, msCCA \cite{wong2020learning}, and stCCA \cite{wong2020inter}. Among these, FBCCA, Compact-CNN, and ensemble DNN are considered SOTA calibration-free approaches.

1) \textbf{FBCCA} 
The FBCCA method involves filtering EEG data using multiple band-pass filters and subsequently calculating the canonical correlation coefficient. The coefficients from various frequency bands are weighted and summed to produce a final result. The FBCCA settings are configured following the guidelines outlined in \cite{chen2015filter}.

2) \textbf{Compact CNN }
Compact CNN shares similarities with EEGNet \cite{lawhern2018eegnet}, and its convolutional architecture enables the automatic extraction of task-relevant SSVEP features.

3) \textbf{ensemble DNN}
The method leverages pre-existing EEG datasets from prior experiments to train a global target identifier DNN, fine-tuned for each participant. This ensemble of fine-tuned DNNs is then applied to a new user. By assessing statistical similarities, the most representative DNNs are selected to predict the target character through a weighted combination. Ensemble DNN is implemented according to https://github.com/osmanberke/Ensemble-of-DNNs.

4) \textbf{eTRCA and TDCA with SAME}
SAME data augmentation uses sine-cosine templates to create artificial signals. These methods are implemented according to https://github.com/RuixinLuo/Source-Aliasing-Matrix-Estimation-DataAugmentation-SAME-SSVEP.

5) \textbf{LST}
This approach seeks to create a versatile transfer-learning framework for enhancing SSVEP-based BCIs by leveraging cross-domain data transfer and is implemented according to https://github.com/TBC-TJU/MetaBCI/blob/master/demos/LST.py.

6) \textbf{msCCA}
msCCA introduces a scheme for learning across multiple stimuli in the target recognition methods, which involves learning data related to both the target stimulus and other stimuli. This approach is implemented according to https://github.com/TBC-TJU/MetaBCI/blob/master/demos/FBMsCCA.py.

7) \textbf{stCCA}
Drawing inspiration from parameter-based and instance-based transfer learning, stCCA method leverages knowledge both within and between subjects, reducing the need for extensive calibration data from a new subject. This approach is implemented according to https://github.com/edwin465/SSVEP-stCCA.

\subsection{Implementation details}
Our approach is implemented using the PyTorch library with a Geforce 3090 GPU, using the Adam optimizer with an initial learning rate of 0.0002. The training epoch is 100, and the batch size is 32.

\subsection{Data Visualization}
\begin{figure} [htbp]
  \begin{subfigure}{0.5\textwidth}
  \centering
    \includegraphics[width=\linewidth]{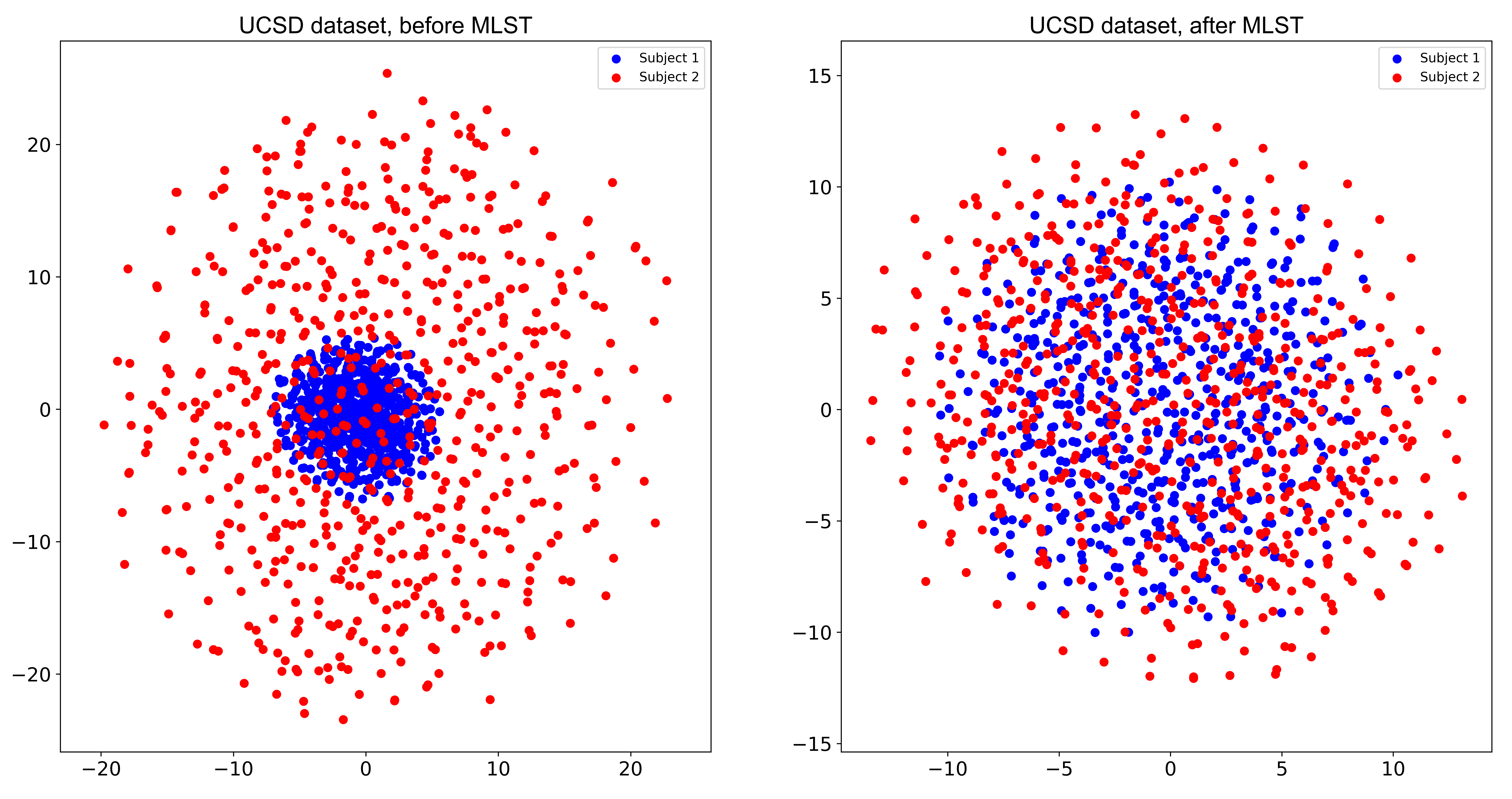}
    \caption{}
  \end{subfigure}
  \begin{subfigure}{0.5\textwidth}
  \centering
    \includegraphics[width=\linewidth]{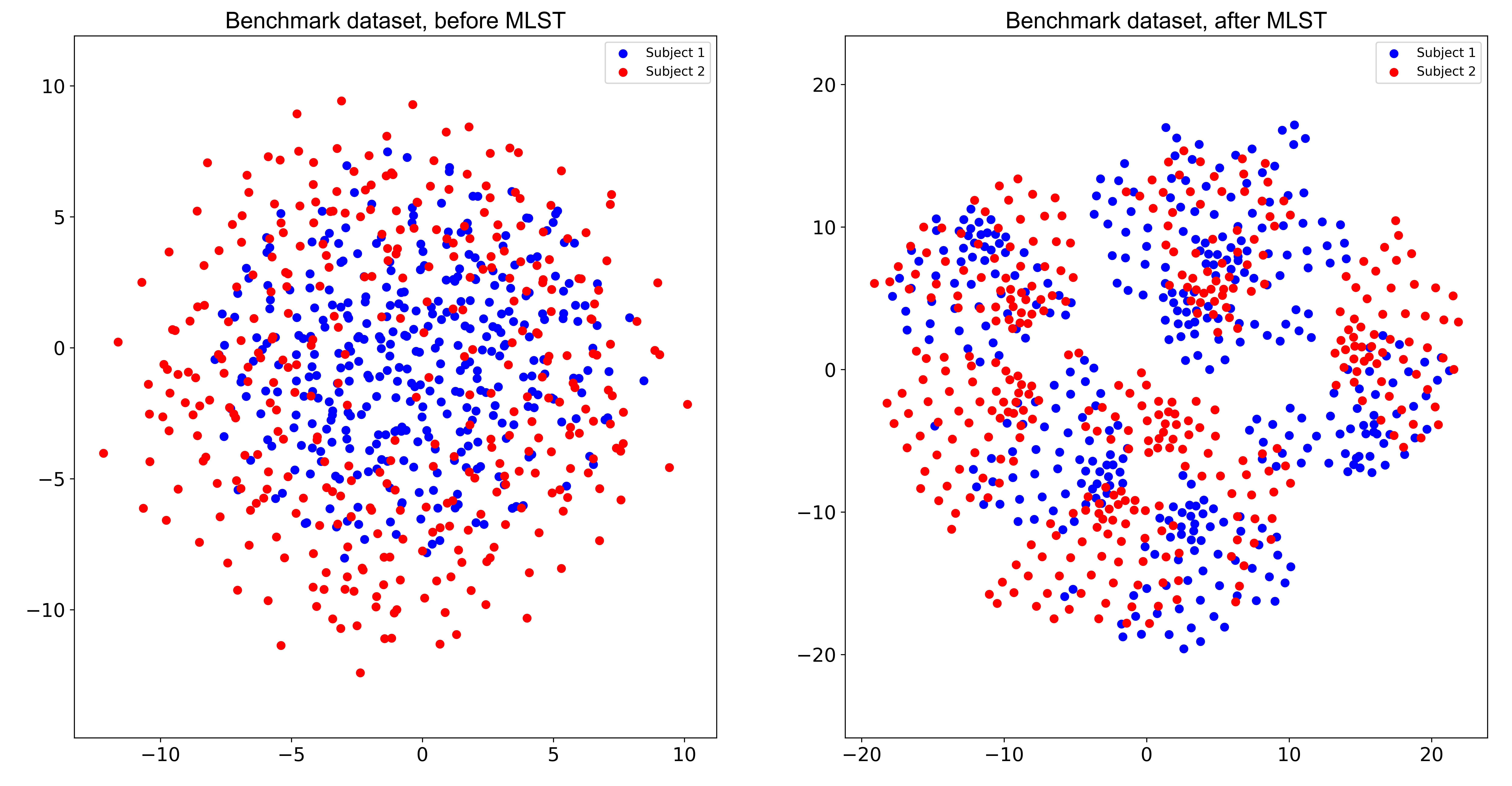}
    \caption{}
  \end{subfigure}
  \begin{subfigure}{0.5\textwidth}
  \centering
    \includegraphics[width=\linewidth]{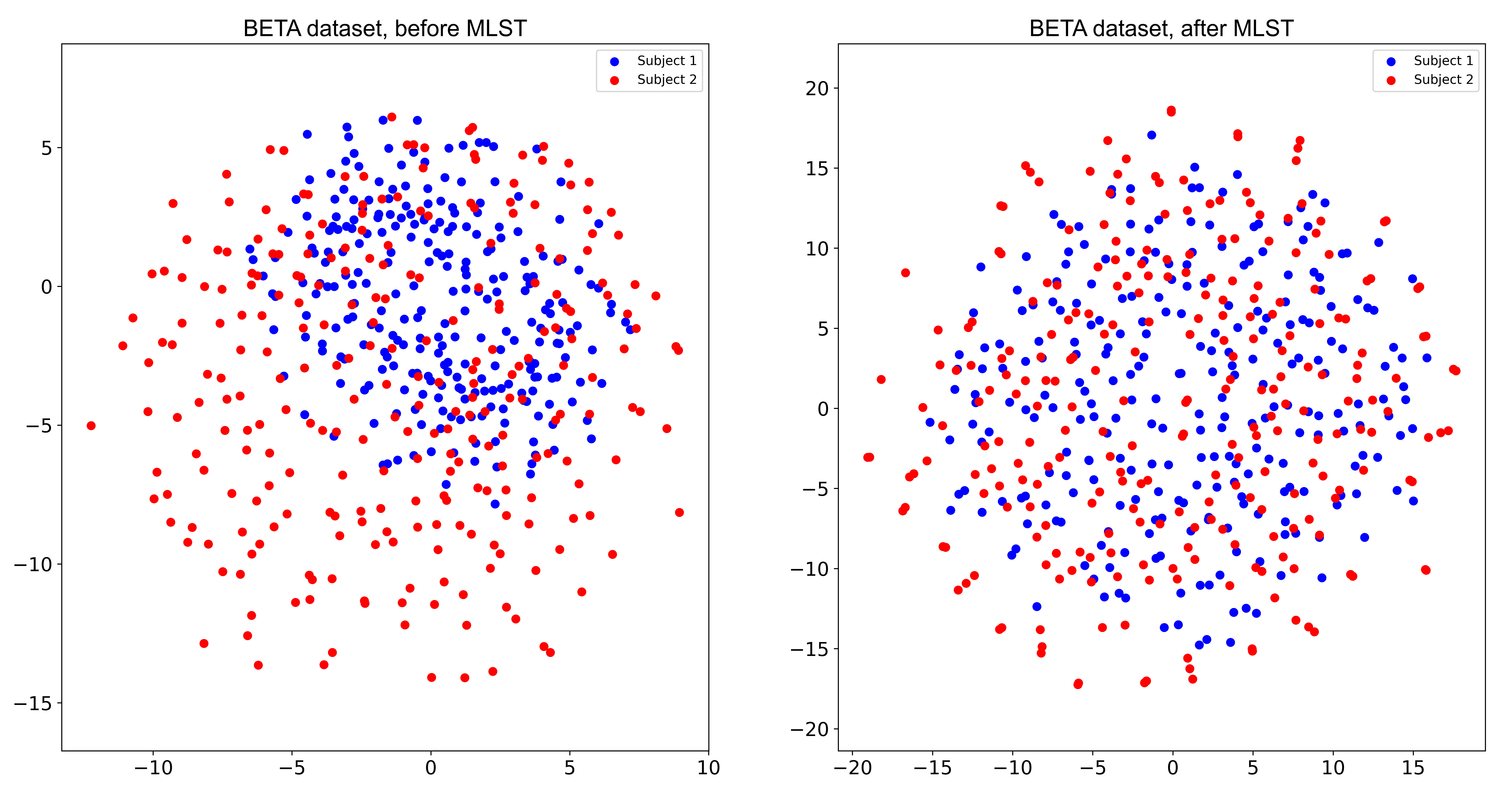}
    \caption{}
  \end{subfigure}

  \caption{\textit{t}-SNE visualization of the data before and after MLST. (a) UCSD dataset; (b) Benchmark dataset; (c) BETA dataset. To get a better look, we have only displayed data from two subjects of one class. }
\label{fig:tsne} 
\end{figure}

\begin{figure*}[h]
\centering
\includegraphics[width=1\textwidth]{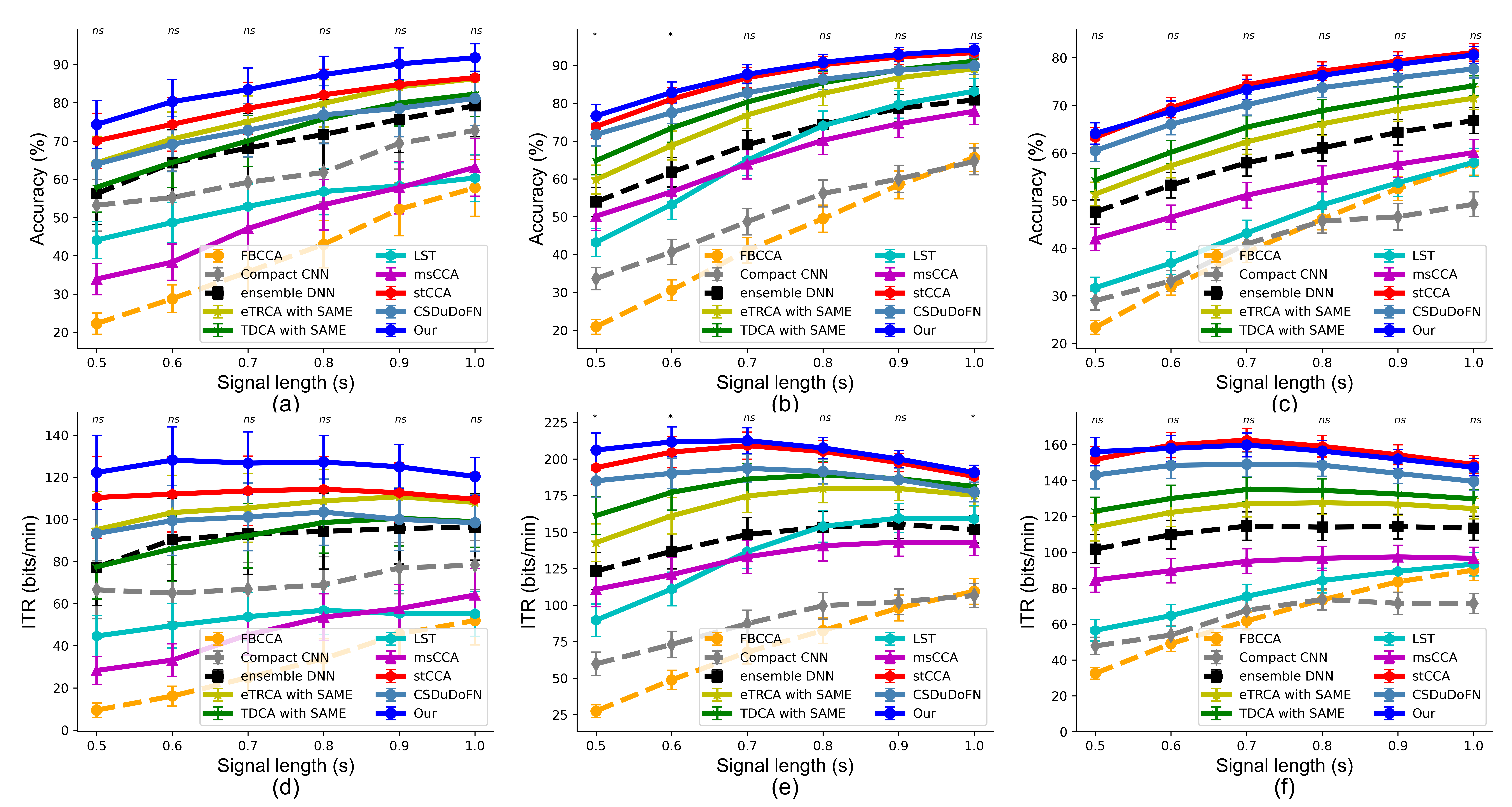}
\caption{Average classification accuracy and ITR are presented for the compared methods on publicly available datasets. Subplots (a)-(c) display accuracy results for the UCSD dataset, the Benchmark dataset, and the BETA dataset, respectively. Subplots (d)-(f) show ITR results for the same datasets. The evaluation is performed using data lengths ranging from 0.5 seconds to 1.0 seconds with 0.1 seconds intervals. In the subfigures, asterisks denote significant differences between stCCA and our proposed method, as determined by paired t-tests (*p \textless 0.05). Error bars represent standard errors.}
\label{fig:result}
\end{figure*}

To demonstrate the effectiveness of MLST in reducing individual differences, we employ t-distributed Stochastic Neighbor Embedding (t-SNE) \cite{Maaten_Hinton_2008}, a nonlinear dimensionality reduction technique used to visualize high-dimensional data in two or three dimensions. We apply t-SNE to compare EEG trials before and after MLST. Fig.~\ref{fig:tsne} (a) presents the t-SNE visualization of two subjects from the UCSD dataset. Blue dots represent trials from subject 1, while red dots represent those from subject 2. For each row, the left plot displays trials before MLST, while the right plot shows trials after MLST. Visualization results for the Benchmark dataset and BETA dataset can be found in Fig.~\ref{fig:tsne}(b) and Fig.~\ref{fig:tsne}(c), respectively. Prior to MLST, the blue and red dots are often widely scattered, indicating a significant separation between subjects. As a result, directly applying transfer learning may yield suboptimal results. However, after MLST, the source and target subjects exhibit overlap, meaning that the differences between them have been reduced. This alignment enhances the potential for successful transfer learning to the target subject.

\subsection{Classification Accuracy on Public Datasets}
The performance evaluated on the three datasets is presented in Fig.~\ref{fig:result}. In the case of the UCSD dataset (refer to Fig.~\ref{fig:result} (a) and (d)), our method outperforms other approaches across all signal lengths. stCCA's highest ITR is 114.31$\pm$15.47 bpm at 0.8 seconds, which is 13.18 bpm lower than ours at 0.6 seconds. Nonetheless, the paired t-test reveals that our method does not exhibit a significant advantage over the stCCA method. This could be attributed to the limited dataset size, which may not satisfy the assumption to use paired t-test that the data follows a normal distribution. It is also possible that our method and the stCCA method have their own strengths and weaknesses, which will be further analyzed in the discussion section. On the Benchmark dataset (refer to Fig.~\ref{fig:result} (b) and (e)), our method consistently outperforms other methods across all signal lengths. Additionally, it exhibits significant differences (p\textless 0.05) in accuracy at 0.5 seconds and 0.6 seconds, as well as in ITR at 0.5 seconds, 0.6 seconds, and 1.0 seconds. Specifically, our method achieves the highest ITR on this dataset, reaching 212.60 at 0.7 seconds, which is 3.44 bpm higher than the stCCA method's maximum ITR of 209.16 at 0.7 seconds. On the BETA dataset (refer to Fig.~\ref{fig:result} (c) and (f)), our method performs slightly worse than the stCCA method except at a signal length of 0.5 seconds. However, this difference is minimal and not statistically significant. Specifically, our method achieves the highest ITR of 159.89, while the stCCA method achieves an ITR of 162.62, both at a signal length of 0.7 seconds. It's worth noting that our proposed CSDuDoFN model overall outperforms the TDCA method with SAME enhancement on the Benchmark and BETA datasets almost across all signal lengths.

\begin{figure}[h]
\centering
\includegraphics[width=0.45\textwidth]{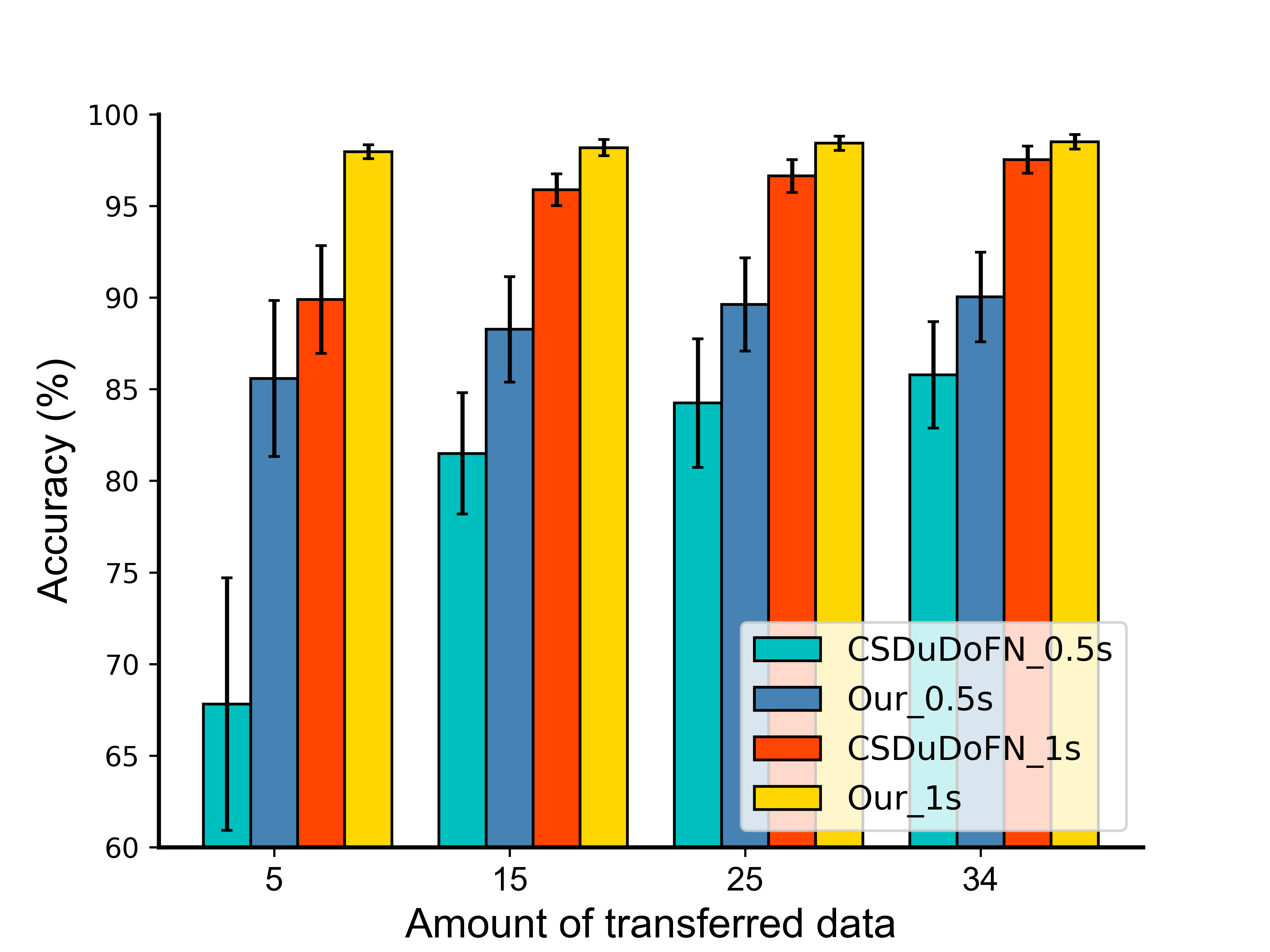}
\caption{Abaltions on the number of data for transfer learning. Error bars represent standard errors.}
\label{fig:transfer}
\end{figure}

\begin{figure*}[h]
\centering
\includegraphics[width=1\textwidth]{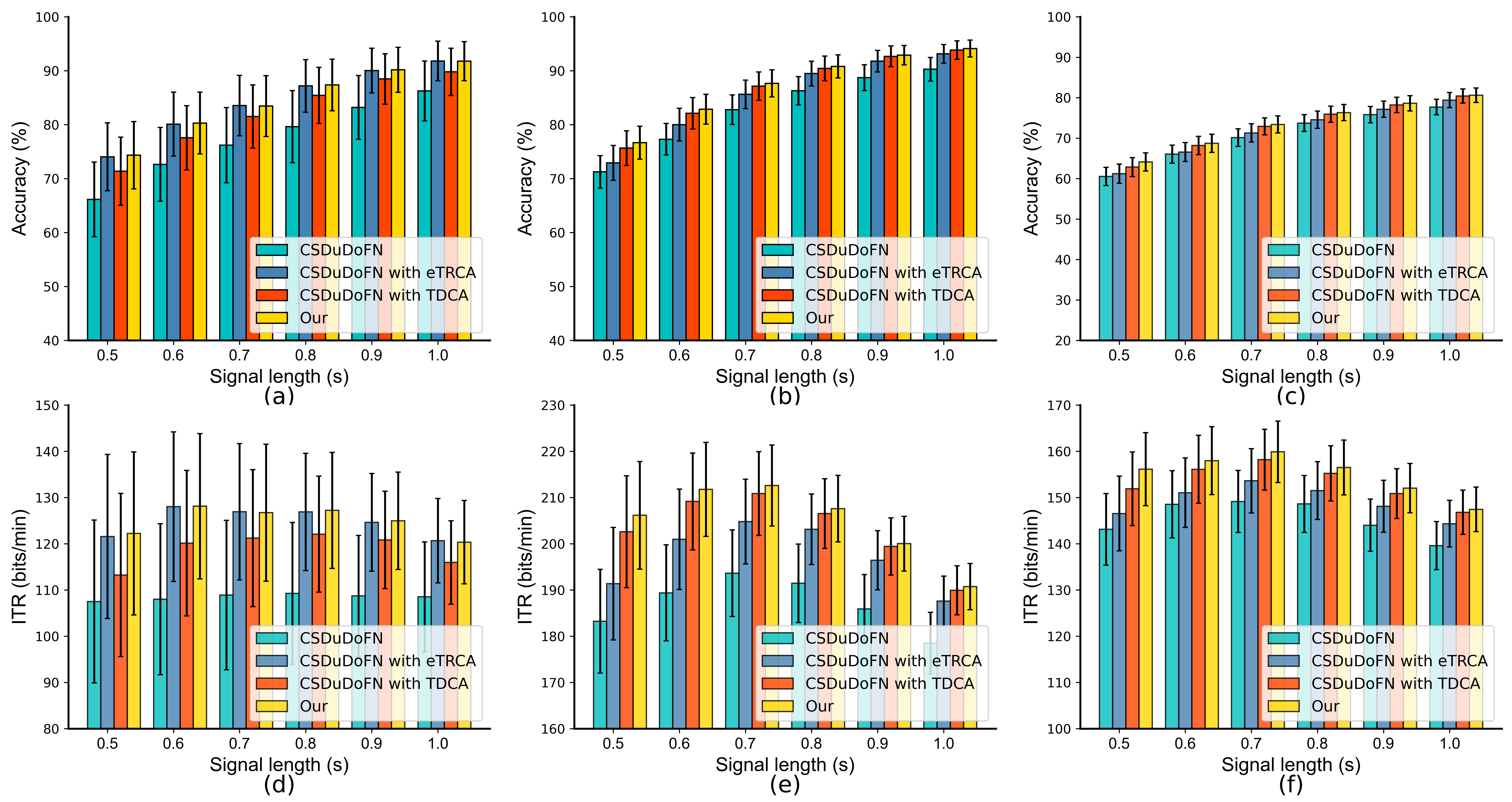}
\caption{Average classification accuracy and ITR are presented for CSDuDoFN, CSDuDoFN with TRCA, CSDuDoFN with TDCA, and ours on publicly available datasets. Subplots (a)-(c) display accuracy results for the UCSD dataset, the Benchmark dataset, and the BETA dataset, respectively. Subplots (d)-(f) show ITR results for the same datasets. The evaluation is performed using data lengths ranging from 0.5 seconds to 1.0 seconds with 0.1 seconds intervals. Error bars represent standard errors.}
\label{fig:ablation}
\end{figure*}

\subsection{Effects of Data for Transfer Learning}
To investigate the effect of data amount on transfer learning performance, we analyze classification accuracies when applying the different amounts of transferred data on the Benchmark dataset. We use subjects 1-6 as testing data and evaluate the accuracy on 0.5 seconds-length and 1.0 seconds-length signals with 5,15,25, and 34 data for transfer learning, separately. As presented in Fig.~\ref{fig:transfer}, we can see that more data used for transfer learning leads to better performance both on the proposed CSDuDoFN model and ours. With a signal length of 0.5 seconds, as the amount of transferred data increases, there is a significant improvement in performance. However, with a signal length of 1.0 seconds, the increase in performance is relatively small.

\subsection{Abaltions on Incorporating eTRCA and TDCA}
Fig.~\ref{fig:ablation} demonstrates the impact of integrating eTRCA and TDCA. Integrating eTRCA and TDCA yields the best performance across all three datasets, while not integrating either eTRCA or TDCA results in the poorest performance. Integrating only TDCA leads to suboptimal performance in the Benchmark dataset and BETA dataset, and integrating only eTRCA results in suboptimal performance in the UCSD dataset. This indicates that TDCA may not necessarily outperform eTRCA in all datasets, and the design of integrating both eTRCA and TDCA is essential for further performance improvement.

\subsection{Abaltions on Dual Domains}
We compare the performance of the proposed CSDuDoFN and our method with and without the use of MLST and original SSVEP signals. Table~\ref{MLST_effect} presents the results for 0.5 seconds and 1.0 seconds data lengths. It can be observed that our method performs optimally when both MLST and original signals are used. When MLST is not used, and only the original signals are employed, our method's performance degrades significantly. Likewise, when only MLST is used, and the original signals are excluded, our method's performance also experiences a moderate decrease. This demonstrates the utility of using both MLST and original signals for improving the method's performance.

\begin{table*}[htbp]
\centering
\begin{tabular}{l|l|l|l|l|l|l}
\hline
\multirow{3}{*}{Methods} & \multicolumn{6}{c}{Datasets} \\
\cline{2-7}
 & \multicolumn{2}{c|}{UCSD} & \multicolumn{2}{c|}{Benchmark} & \multicolumn{2}{c}{BETA} \\
\cline{2-7} & $0.5 \mathrm{~s}$ & $1.0 \mathrm{~s}$ & $0.5 \mathrm{~s}$ & $1.0 \mathrm{~s}$& $0.5 \mathrm{~s}$ & $1.0 \mathrm{~s}$ \\\hline
CSDuDoFN w/o MLST & $43.28\pm5.60$ & 58.61$\pm$6.88 & $40.48\pm3.48$ & $69.23\pm3.78$&$36.54\pm2.18$ & $56.18\pm2.78$ \\
CSDuDoFN w/o original SSVEP & $\mathbf{65.23\pm5.19}$ & $\mathbf{86.24\pm4.91}$ & $70.54\pm3.16$ & $\mathbf{91.49\pm1.91}$&$56.70\pm2.41$ & $77.51\pm1.94$ \\
CSDuDoFN & $63.96\pm6.75$ & $81.16\pm7.09$ & $\mathbf{71.72\pm3.00}$ & $89.94\pm2.29$&$\mathbf{60.53\pm2.22}$ & $\mathbf{77.69\pm1.93}$ \\\hline
\hline Our w/o MLST & $67.51\pm6.81$ & $86.37\pm5.73$ & $69.36\pm3.33$ & $91.92\pm2.06$&$58.98\pm2.33$ & $77.29\pm1.90$ \\
Our w/o original SSVEP & $74.04\pm5.42$ & $91.47\pm3.59$ & $75.44\pm3.18$ & $94.11\pm1.54$&$62.48\pm2.36$ & $80.34\pm1.80$ \\
Our & $\mathbf{74.27\pm6.08}$ & $\mathbf{91.77\pm3.63}$ & $\mathbf{76.65\pm3.06}$ & $\mathbf{94.12\pm1.57}$&$\mathbf{64.11\pm2.25}$ & $\mathbf{80.61\pm1.78}$ \\
\hline
\end{tabular}
\caption{Comparison of the accuracy of CSDuDoFN with and without MLST or original SSVEP and our proposed approach with and without MLST or original SSVEP, conducted at signal lengths of 0.5 seconds and 1.0 seconds across different datasets.}
\label{MLST_effect}
\end{table*}

\section{Discussion}\label{sec:Dis}
\subsection{Comparison with One-shot Learning in Computer Vision}
Taking a closer look at our proposed approach, it aims to achieve two primary objectives: i) minimize the domain differences between the source and target data to facilitate better transfer learning. We employ the MLST operation to achieve this; ii) maximize the utilization of target data. We accomplish this using the SAME data augmentation \cite{luo2022data}. Our approach shares similarities with a prevalent strategy employed in current practices for few-shot learning \cite{wang2020generalizing}, one-shot learning \cite{vinyals2016matching} or meta-learning \cite{hospedales2021meta} in the field of computer vision. This strategy entails the training of a base model with additional data to harness the full potential of existing data. Subsequently, fine-tuning is performed on the support set to maximize the utilization of target data. However, SSVEP signals exhibit substantial inter-subject variability, and in contrast to natural images, they lack common features such as texture, shape, and color information. Furthermore, the limited size of SSVEP datasets, compared to their image counterparts, renders fine-tuning with only one calibration data ineffective (as evidenced by our unsuccessful attempts). Consequently, our chosen strategy involves augmenting the one-shot data and integrating it with the network's output.

\subsection{Comparison with stCCA}
Traditional calibration-based SSVEP classification schemes primarily focus on acquiring subject-specific and class-specific parameters, which encompass spatial filters and SSVEP templates. In contrast, the stCCA method adopts a multi-frequency approach to capture subject-specific information. Furthermore, it leverages a weighted summation of spatially filtered SSVEP templates from source subjects to approximate the spatially filtered SSVEP template of the target subject, facilitating the extraction of class-specific information. Diverging from the stCCA method, our approach takes a distinct route when dealing with the acquisition of class-specific parameters for the target data. Instead of computing the weighted sum of filtered source data, we directly project both the source domain data and the target domain data into a unified sine-cosine template domain. This approach offers heightened interpretability compared to stCCA. On the other hand, for learning the subject-specific parameters of the target data, we employ a data augmentation operation on the source data. In terms of performance, our approach exhibits superior accuracy on the UCSD dataset compared to the stCCA method, although without significant differences. Conversely, on the BETA dataset, the stCCA method demonstrates overall better performance than our method, also without significant differences. This suggests a potential complementarity between our method and the stCCA method. To further explore this, we examine the specific performance of these two methods on the UCSD dataset and observe that their effectiveness varies across different subjects. One method may excel with certain subjects but fare less favorably with others, as illustrated in Table~\ref{difference}.
 
\begin{table}[htbp]
\begin{tabular}{l|cccc}
\hline
\diagbox{Subject}{Accuracy (\%)}{Method} & stCCA & Our& stCCA & Our \\
\hline S01 & 26.35 & \textbf{42.02}& 40.04 & \textbf{72.54} \\
S02 & \textbf{53.65} & 43.45& \textbf{73.13} & 66.83 \\ 
S03 & 44.93 & \textbf{70.00}& 71.98 & \textbf{92.30} \\ 
S04 & 80.20 & \textbf{81.27}& \textbf{99.09} & 98.73 \\ 
S05 & \textbf{98.29} & 97.06& \textbf{100.00} & 99.96 \\ 
S06 & 81.31 & \textbf{92.54}& 96.71 & \textbf{99.48} \\ 
S07 & 83.82 & \textbf{86.30}& 96.39 & \textbf{97.42} \\ 
S08 & \textbf{94.17} & 90.75& 99.60 & \textbf{99.88} \\ 
S09 & 74.72 & \textbf{82.58}& 96.35 & \textbf{99.13} \\ 
S10 & \textbf{64.21} & 56.75& \textbf{92.34} & 91.39 \\\hline
Average & 70.06 & \textbf{74.27}& 86.56 & \textbf{91.77} \\\hline

\end{tabular}
\caption{Comparison between the accuracy of the stCCA and our proposed approach for different subjects in the UCSD dataset at 0.5 seconds and 1.0 seconds. The first two columns correspond to the results at 0.5 seconds, while the last two columns correspond to the results at 1.0 seconds}
\label{difference}
\end{table}

\subsection{Limitation and Future Work}
As our method requires a target subject to have one calibration data for each stimulus, it can only be used when $K$=$N_f$. In the future, we aim to extend our approach to scenarios where $K$ \textless $N_f$, using a methodology similar to multi-frequency learning\cite{wong2020learning}. Our approach employs sine-cosine reference signals as templates. However, this may not be optimal for SSVEP recognition due to the inter-subject variability of SSVEP and the influence of ongoing EEG and noise \cite{zhang2011multiway}, as well as the absence of important information contained in real EEG data \cite{zhang2014frequency}. Therefore, in the future, we plan to investigate the incorporation of personalized regularization terms when constructing reference signals. Additionally, we will explore the integration of our method with the stCCA approach to leverage their complementary strengths and further enhance classification performance. We also intend to investigate how to extend our approach to scenarios without calibration data, developing high-performance BCI systems that do not require calibration.

\section{Conclusion}\label{sec:Con}
This study introduces an innovative method for one-shot SSVEP classification. We have introduced the MLST to project source and target data into the sine-cosine domain, effectively reducing inter-subject variability. Additionally, our novel CSDuDoFN model capitalizes on original SSVEP signals and their transformed counterparts. The final classification is achieved by combining the output of the CSDuDoFN with SAME-augmented eTRCA and TDCA. The effectiveness of our proposed approach is rigorously evaluated across three distinct SSVEP datasets, yielding superior performance on two datasets and remaining highly competitive on the third. This research underscores the potential for seamless integration of BCIs into everyday life.

\ifCLASSOPTIONcaptionsoff
  \newpage
\fi

\bibliographystyle{IEEEtran}
\bibliography{ref}
\end{document}